\documentclass[preprint,11pt]{JHEP3} 

\usepackage{epsfig,multicol,amsmath}

\def\beq{\begin{equation}}
\def\eeq{\end{equation}}
\def\bea{\begin{eqnarray}}
\def\eea{\end{eqnarray}}
\def\pom{{I\!\!P}}
\def\eq#1{{Eq.~(\ref{#1})}}
\def\fig#1{{Fig.~\ref{#1}}}

\newcommand{\nn}{\nonumber}

\newcommand{\Lb}{\left(}
\newcommand{\Rb}{\right)}

\newcommand{\p}{I\!\!P}

\parskip=\itemsep               

\newcommand{\h}{\frac{1}{2}}

\title{\LARGE \bf  Inelastic processes in DIS  and    N=4 SYM }
\author{\large  E. ~Levin$^{a,b}$\,\, and \,\,I.~Potashnikova$^{a}$  \\
a)\,Departamento de F\'\i sica, Universidad T\'ecnica,
Federico Santa Mar\'\i a, Avda. Espa\~na 1680,
Casilla 110-V, Valpara\'\i so, Chile \\
b)  \,Department of Particle Physics, School of Physics and Astronomy,
Raymond and Beverly Sackler
 Faculty
of Exact Science,  Tel Aviv University, Tel Aviv, 69978, Israel}

\abstract{In this paper we compare the prediction for deep inelastic scattering from N=4 SYM with the HERA experimental data. The paper conveys two results.  The first is the message that N=4 SYM is able to describe the DIS data with very good accuracy ($\chi^2/d.o.f. \,\leq\,1.5$) in the region of $Q^2 = 0.85 \div  60\,GeV^2$ with $2/\sqrt{\lambda} = 0.7 \div 0.8 $ . The second is that the value of string coupling constant  $g_s$ turns out to be so small that none of saturation effects will be visible in the region of accessible energies,   including the maximal energy of the LHC (W = 14 \,\,TeV). }


\preprint{ 
\today}
\begin{document}

\numberwithin{equation}{section}

\section{Introduction}
It is well known that  N=4  SYM  together with AdS/CFT correspondence
 allows us to study theoretically the regime of the strong coupling constant \cite{AdS-CFT} . For the first time we have a theory  which leads to the main ingredients of  the high energy phenomenology such as the Pomeron and the  Reggeons, in 
 the limit of strong coupling.  On the other hand, N=4 SYM with small coupling leads to normal QCD like physics (see Refs.
\cite{POST,BFKL4})  with OPE  and linear equations for DIS as well as the BFKL equation for the high energy amplitude.

The Pomeron which appears in N=4 SYM\cite{BST1} has the intercept and the slope of the trajectory  that are equal to
\beq\label{I1}
\alpha_{\p}\Lb 0 \Rb\,\,\,=\,\,\,2\,\,-\,\,\frac{2}{\sqrt{\lambda}}\,\,\equiv\,\,\,\,2 \,-\,\rho;\,\,\,\,\,\,\,\,\,\,
\alpha'_{\p}\Lb 0 \Rb\,\,\,=\,\,\,0.
\eeq
in the limit of $\rho \,\ll\,1$. First, we would like to recall that  N=4 SYM has a simple solution  for the following set of couplings:
\beq \label{I2}
g_s\,\,=\,\,\frac{g^2_{YM}}{4 \pi}\,\,=\,\,\alpha_{YM}\,\,
\,=\,\,\frac{\lambda}{4 \pi N_c};\,\,\,\,\,\,\,R\,\,=\,\, \alpha'^{\frac{1}{2}}\,\lambda^{\frac{1}{4}};\,\,\,\,\,\,\,\,g_s\,\,\ll\,\,1;\,\,\,\,\mbox{but}\,\,\,\,\,\lambda\,\,\gg\,\,1
\eeq
where $R$ is the radius in  $AdS_5$- metric:
\beq \label{I3}
d s^2\,\,=\,\,\frac{R^2}{z^2}\,\Lb \,d z^2\,\,+\,\,\sum^d_{i=1}  d x^2_i \Rb\,=\,\frac{R^2}{z^2}\,\Lb \,d z^2\,+\,dx_\mu dx^\mu \Rb
\eeq
with $\mu = 0,1,2,3$.

One can see that at large $\lambda $ the Pomeron intercept is close to 2 and, therefore, the exchange of the Pomeron gives almost real amplitude. Indeed, the unitarity constraint in this case looks as follows  \cite{MHI,BST1,BST2,BST3,COCO,BEPI,LMKS}
\beq \label{I4}
\mbox{Im}\,A\Lb s, b; z,z'\Rb\,\,\,=\,\,\,|A\Lb s, b; z,z'\Rb|^2\,\,\,\,+\,\,\,\,{\cal O}\Lb \rho = \frac{2}{\sqrt{\lambda}}\Rb
\eeq
\eq{I4} means that the contribution of the multiparticle production is small  for the strong coupling and main source of the total cross section is originated by elastic and quasi-elastic (
diffractive)  processes when the target (proton) remains
intact. 
Such a picture not  only contradicts
the QCD expectations\cite{GLR,MUQI,MV,BFKL,BK,JIMWLK}, but also contradicts available experimental data.

On the other hand, the main success of N=4 SYM  has been achieved in the description of the multiparticle system such as quark-qluon plasma and/or the multiparticle system at fixed temperature \cite{KSS,HKKKY,MUN4,PLASMAN4}. 

Therefore, we have either to  find a new mechanism for multiparticle production in N=4 SYM
(see an attempt in Ref.\cite{KHLE}) or to assume that $\lambda$ is not very large (say $\rho = 0.5 \div 0.8$). 
It should be noticed that even at $\rho =0.8$ $\lambda$ is rather large $\approx 6$.

The first goal of this paper is to find the range of $\lambda$ that can describe the deep inelastic(DIS)  data from HERA.  We believe that correction to \eq{I1} is proportional to $1/\lambda$ and $\lambda  \approx 6$
could lead to a good   in the description of the experimental data. It  can give  a sizable cross section for the multiparticle production. Such an approach will be a N=4 SYM
motivated model which will be able to  provide a guide for a theoretical  approach to QCD in the region of strong coupling.

It was shown  in Refs.\cite{MHI,BST2,BST3,COCO,BEPI,LMKS} that we face the saturation  phenomena  in N=4 SYM at low $x$ for DIS.  The physics of the saturation looks very similar to the saturation phenomena  in high density QCD with one essential difference:
  the saturation in N=4  SYM we can theoretically describe in very simple fashion based on the eikonal formulae. This approach  can be easily generalized for scattering both dense and diluted systems. Using these formulae we can learn what can happen in  the region of very low photon virtualities and find parameters such as $g_s$ that govern the strong interactions.

The second goal of this paper is to find parameters of N = 4 SYM that characterize the strength of interaction in the large coupling limit from  the comparison with DIS data and derive the estimates for the expected saturation effects.

N=4 SYM  being conformal invariant theory has only massless particles and leads to the scattering amplitudes that fall as a power of $b$  at large values of impact parameters ($b$). This decrease results in the power- like dependance of typical impact parameters 
in the amplitude. In particular,  for hadron-hadron scattering these typical $b \,\,\propto s^{1/3}$\cite{LMKS,LEPO} which contradicts the Froissart theorem \cite{FROI}.  We have to go beyond of N=4 SYM  and discuss the string theory, conformal limit of which is N=4 SYM, to restore the logarithmic behaviour ( $ b  \propto  \ln s$)  of the typical impact parameters.
The third goal of this paper is to find out how the $b$ behaviour influences the description of the experimental data.

In the paper we compare the N=4 SYM formula for deep inelastic structure function $F_2$ which we derive in the next section, with the HERA data for low $x$ region ($ x \leq 0.01$).  Since the physical meaning of  the fifth coordinate $z$  (see \eq{I3}) is the typical size of the colliding particles, DIS gives an unique opportunity to check the predicted behaviour on $z$. On the other hand, it is known that the energy dependance of DIS is rich and, in particular, $F_2\,\propto \, x^{ - \lambda}$ with $\lambda \,=\,0.1 \div 0.5$ for $Q^2 = 0.1 \div 27 GeV^2$, respectively. Therefore, we have 
the set of the experimental data both for checking energy and $z$ dependance of the scattering amplitude, especially because the
experimental errors are so small that it is a challenge to describe the data in any theoretical approach (see section 3 of this paper).

The paper conveys two results.  The first is the message that N=4 SYM is able to describe the DIS data with very good accuracy ($\chi^2/d.o.f. \,\leq\,1.5$) in the region of $Q^2 = 0.85 \div  60\,GeV^2$ with $\rho = 0.7 \div 0.8 $   ( see section 3). The second is that the value of $g_s$ turns out to be so small that none of saturation effects will be visible in the region of accessible energies  including the maximal energy of the LHC (W = 14 \,\,TeV).
DIS data can be described both  in conformal N=4 SYM and  taking into account non-conformal corrections in $b$ dependance.

However for description of proton-proton scattering we need corrected $b$ dependance (see section 3). The main result of Ref. \cite{LEPO} that it  should be the other source of the multiparticle production than N=4 SYM remains even for $\rho = 0.7 \div 0.8$.

\section{High energy Scattering in N=4 SYM}
\subsection{Pomeron exchange}
As has been mentioned there exists the Pomeron in N=4 SYM with the parameters of its trajectory given by \eq{I1}. The exchange of this Pomeron leads to the following contribution to the scattering amplitude (see \fig{pom}-a):
\beq \label{HES1}
\tilde{A}\,\,=\,\,\frac{g^2_s}{4 \pi}\,\Big\{\frac{2}{\pi \rho}\,+\,i\Big\}\,\,\frac{\Lb  z_1 z_2 s\Rb^{1 - \rho}}{\sqrt{u\,(2 + u)}}\,\,\,\,\frac{\ln\Lb 1 + u + \sqrt{u\, (2 + u)}\Rb}{\sqrt{\rho\, \,\pi \,\ln^3\Lb  z_1 z_2 s\Rb}}\,\exp\Lb - \frac{\ln^2\Lb 1 + u + \sqrt{u (2 + u)}\Rb}{\rho 
\ln\Lb z_1 z_2 s\Rb}\Rb
\eeq
where
\beq \label{U}
u\,=\,\frac{( z_1 - z_2)^2 + b^2}{2 z_1 z_2}\,\,\,\,\,\,\mbox{and}\,\,\,b \,\,\mbox{is the impact parameter in the scattering amplitude}
\eeq

\FIGURE[h]{\begin{tabular}{c c}
\epsfig{file=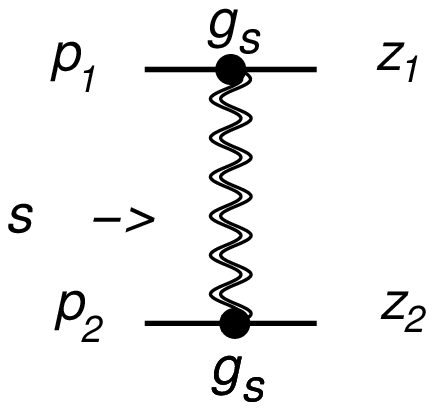,width=25mm}& \epsfig{file=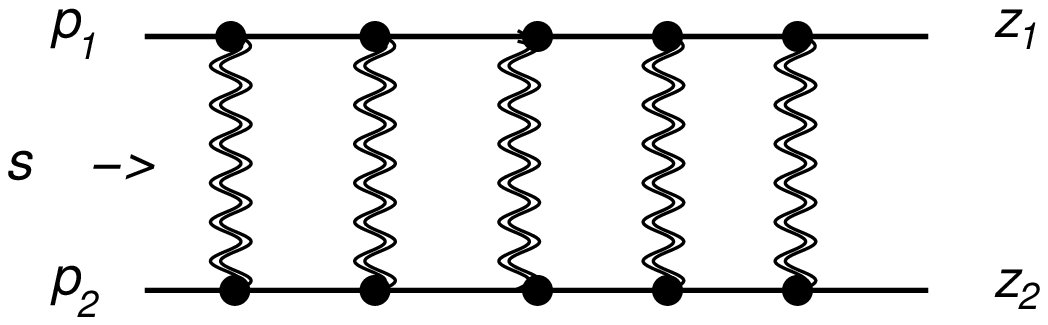,width=65mm}\\
\fig{pom}-a & \fig{pom}-b\\
\end{tabular}
\caption{It is shown the one  Pomeron ( reggeized graviton)  exchange in \fig{pom} -a and the eikonal rescattering (\fig{pom}-b) for N=4 SYM at large coupling}
\label{pom}}

One can see that \eq{HES1} is very similar to the expression for the exchange of the BFKL Pomeron\cite{BFKL}  in which
the sizes of the interacting dipoles are replaced by $z_1$ and $z_2$ and in which
 $\Delta_\pom = 2 - \alpha_\pom(0)$ and 
the diffusion coefficient are  equal. It should be recalled
that \eq{HES1} describes  high energy scattering in the kinematic region  where $z_1 z_2s \,\gg\,\lambda \,\gg\,1$.
\subsection{Eikonal formula}
 Since the Pomeron intercept is larger than 1,  considering the high energy scattering  we cannot restrict ourselves by the exchange of one Pomeron.  It is well known that in N=4  at small coupling as well as in  perturbative  QCD the problem to take into account all Pomeron exchanges and the Pomeron interaction is a very difficult problem that has been only partly solved in high density QCD (see Refs. 
 \cite{GLR, MUQI,MV,BFKL,BK,JIMWLK}). However, for N=4 SYM with large  coupling the situation turns out to be much simple and at small values of $\rho$ the amplitude can be found in the eikonal approximation\cite{BST2,BST3,COCO,LMKS} (see \fig{pom}-b), namely,

\bea \label{HES3}
&&A\Lb s, b; z_1, z_2\Rb\,\,=\,\,i \Big\{ 1\,\,-\,\,\exp\Lb i \tilde{A}\Lb s,b;z_1,z_2\Rb\Rb\Big\}\nonumber
\eea
The total cross section that we are going to discuss is proportional to imaginary part of the amplitude and can be written in the form for virtual photon- proton scattering in the form
\bea \label{HES4}
&&\sigma_{tot}\Lb \gamma^* + p \Rb\,\,=\\
&&\,\,2 \int d^2 b \int^\infty_0  d z_1 d z_2 \Phi_{\gamma^*}\Lb z_1\Rb\,\Phi_{\mbox{proton}}\Lb z_2\Rb\,\,\Big\{1\,\,-\,\,cos\Lb N^2_c\mbox{Re}\tilde{A}\Lb s,b;z_1,z_2\Rb\Rb\,\exp\Lb - N^2_c\mbox{Im}\tilde{A}\Lb s,b;z_1,z_2\Rb\Rb\Big\}\nonumber
\eea
Functions $ \Phi_{\gamma^*}\Lb z_1\Rb$ and $\Phi_{\mbox{proton}}\Lb z_2\Rb$ describe the probability   
for virtual photon and proton to have the size $z_1$ and $z_2$, respectively, and we will discussed them in the next section.

\begin{boldmath}

\subsection{ $\Phi_{\gamma^*}$ and  $\Phi_{\mbox{proton}}$}

\end{boldmath}

At low $x$ the DIS on the boundary can be expressed through the dipole -proton cross section 
\beq \label{FPHI1}
\sigma_{tot}\Lb \gamma^* p\Rb \,\,=\,\,\int d^2 r_\perp P_{\gamma^*}\Lb r_\perp \Rb\,\,\sigma_{tot}\Lb \mbox{dipole-proton}; r_\perp;x\Rb
\eeq
where the probability to find a dipole with the size $r_\perp$ $P_{\gamma^*}\Lb r_\perp \Rb$ is equal to \cite{WFPH}
\beq \label{FPHI2}
P_{\gamma^*}\Lb r_\perp \Rb \,\,=\,\,\frac{\alpha_{em} N_c}{2 \pi^2 }\,\sum^{N_f}_1 Z^2_f [\zeta^2 + (1 - \zeta)^2]\,\bar{Q}^2\,K^2_1\Lb \bar{Q} r_\perp \Rb
\eeq

where $\zeta$ is the fraction of the energy that is carried by the quark,   $Z_f$ is the fraction of the electric charge for the quark of flavour $f$,  $\alpha_{em}$ is the electromagnetic fine constant.

To find $\Phi_{\gamma^*} $   we need to generalize the wave function of the photon ($K_0\Lb \bar{Q} r_\perp \Rb$)
on the boundary to the wave function in the bulk.
 We can reconstruct  this wave function
 using the Witten formula \cite{WIT}, namely,

\bea \label{FPHI3}
&& \Psi_{\gamma^*} \Lb r,  z\Rb\,\,\,=\\
&&\frac{\Gamma\Lb \Delta\Rb}{\pi\,\Gamma\Lb \Delta - 1\Rb}\,\,\int \,d^2 r' \,
\Lb \frac{z}{z^2\,\,+\,\,( \vec{r}\,-\,\vec{r}')^2}\Rb^{\Delta}\,\, \Psi_{\gamma^*}\Lb r'_\perp\Rb
\,\,\,\mbox{with}\,\,\,\,\Delta_{\pm}\,\,=\,\,\h \Lb d\,\,\pm\,\,\sqrt{d^2 + 4\, m^2}\Rb \nonumber
\eea
where $\Psi\Lb r´_\perp\Rb$ is the wave function of the dipole inside the photon on the boundary. Using \eq{FPHI3} we can find $\Phi_{\gamma^*}(z) $ as
\bea \label{FPHI4}
 \Phi_{\gamma^*}\,\,& = &\,\frac{\alpha_{em} N_c}{2 \pi^2 }\,\sum^{N_f}_1 Z^2_f \int^1_0 d \zeta [\zeta^2 + (1 - \zeta)^2]\,\bar{Q}^2\, \int d^2 r  |\Psi_{\gamma^*} \Lb r,  z\Rb|^2\,\,= \\
& = &\,\,\frac{\alpha_{em} N_c}{2 \pi^2 }\,\int^1_0 d \zeta \sum^{N_f}_1 Z^2_f [\zeta^2 + (1 - \zeta)^2]\,\bar{Q}\, \int d^2 r
d^2 r^,\,d^2 r^{,,}\Lb\frac{\Gamma\Lb \Delta\Rb}{\pi\,\Gamma\Lb \Delta - 1\Rb}\Rb^2 \nn\\
&& \times \Lb\frac{z}{z^2\,\,+\,\,( \vec{r}\,-\,\vec{r}^,)^2}\Rb^{\Delta}\,\, K_1\Lb \bar{Q} r^,\perp\Rb\,\,\Lb\frac{z}{z^2\,\,+\,\,( \vec{r}\,-\,\vec{r^{,,}})^2}\Rb^{\Delta}\,\, K_1\Lb \bar{Q} r^{,,}_\perp\Rb\nn
\eea

Using the formulae {\bf 3.198},  {\bf 6.532(4)}, {\bf 6.565(4)}  and {\bf 6.566(2)} from the Gradstein and Ryzhik Tables, Ref. \cite{RY} we can rewrite \eq{FPHI4} introducing the Feynman parameter ($\xi$) and taking the integral over $ r $ and the angle between $\vec{r}^,$ and $\vec{r}^{,,}$. It has the form
\bea \label{FPHI5}
 \Phi_{\gamma^*}\,\,& = &\,z^4\frac{\alpha_{em} N_c}{2  }\,\sum^{N_f}_1 Z^2_f \int^1_0 d \zeta\, [\zeta^2 + (1 - \zeta)^2]\,\bar{Q}^2\, \int d r^{, 2}  d r^{,, 2} \int^1_0 d \xi\,K_1\Lb \bar{Q} r^,_\perp\Rb K_1\Lb \bar{Q} r^{,,}_\perp\Rb \\
 &\times& \,\,\,\frac{2 \Lb \xi (1 - \xi)(r^{, 2} + r^{,,  2}) + z^2 \Rb^2   +  4 r^{,} r^{,,}\xi^2 (1 - \xi)^2 }{
 \Lb \xi (1 - \xi)( r^, - r^{,,})^2  + z^2 \Rb^{5/2}\, \Lb \xi (1 - \xi)( r^, + r^{,,})^2  + z^2 \Rb^{5/2}}\nn
\eea
In \eq{FPHI5} we used that for photon $m$ and $d$  in \eq{FPHI3} are equal to 0 and 2, respectively.
\FIGURE[h]{
  \includegraphics[width=8 cm] {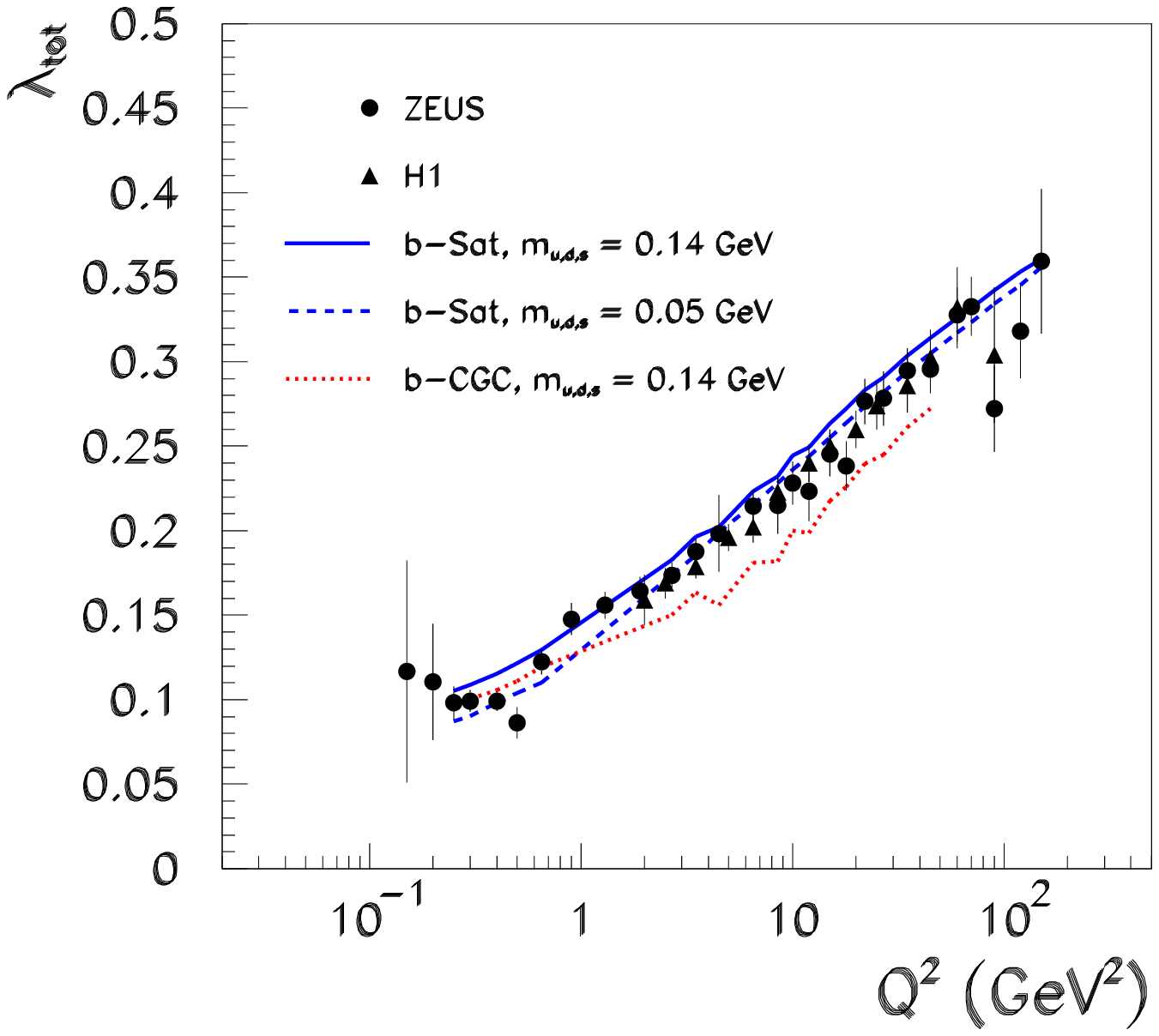}
\caption{The slope $d \ln F_2/d \ln(1/x)$ versus $Q^2$. Data are taken from Ref.\cite{KMW}. In this paper the slope was extracted from the data of Ref.\cite{H1ZEUSlam}. The curves are the fit to the data, based on perturbative QCD,  given in Ref.\cite{KMW}.}
\label{slopef2}
}

~

~

For   $\Phi_{\mbox{proton}}$ we use the expression that has been suggested in Ref. \cite{LEPO}, namely,

\beq \label{FPHI6}
\Phi_{\mbox{proton}}\Lb z\Rb\,\,=\,\,\int \, d^2 r \prod^{N_c}_{i =1}\,|\Psi\Lb r_i,z\Rb|^2\,\,
\eeq
In \eq{FPHI6} we assumed that a proton consists of $N_c$ colourless dipoles and each dipole interacts with other dipoles without correlation.
We use \eq{FPHI3} to find out function $\Psi\Lb r_i,z\Rb$ .
In this equation  $\Psi\Lb r'\Rb$ is the wave function of the dipole inside the proton on the boundary. For
simplicity and to make all calculations more transparent, we choose  $\Psi\Lb r'\Rb = K_0\Lb q r'\Rb$.
The value of the parameter $Q$ can be  found from the value of  the electromagnetic radius of the proton ($q\,\approx 0.35\,GeV^{-1}$).

Substituting \eq{FPHI3} in \eq{FPHI6} one obtains

\bea \label{FPHI7}
&&\Phi_{\mbox{proton}}\Lb z\Rb\,\,=
\,\,\frac{N_c}{\cal N}  \,\,\Lb\frac{\Gamma\Lb \Delta\Rb}{\pi\,\Gamma\Lb \Delta - 1\Rb}\Rb^2\,\,\nn\\
&&\times\,\,\,\int \,d^2 r^, \,d^2 r^{,,}\,d^2 r\, \,K_0\Lb q r'_\perp\Rb\, \,K_0\Lb q r^{,,}_\perp\Rb\\
 &&\times\,\,\,
\Lb \frac{z}{z^2\,\,+\,\,( \vec{r}\,-\,\vec{r}')^2}\Rb^{\Delta}\,\,
\Lb \frac{z}{z^2\,\,+\,\,( \vec{r}\,-\,\vec{r}^{,,})^2}\Rb^{\Delta}\nn
\eea
where $\cal{N}$ is the norm of the dipole wave function on the boundary ( ${\cal{N}} =\pi/ q^2$ for $K_0\Lb q r_\perp\Rb$)

Using that
\bea\label{FPHI8}
&&\int^{\infty}_0  \frac{J_0\Lb k r \Rb\,r\,d r }{\Lb z^2 + r^2\Rb^{\Delta}} \,  = \, \frac{1}{\Gamma\Lb \Delta\Rb}
2^{1 - \Delta}\,k^{\Delta - 1}\, K_{1 -\Delta }\Lb k r\Rb;\,\,\,\,\,\nn\\
&&
\int^{\infty}_0  r d r \, J_0\Lb k r \Rb\, K_0\Lb q r\Rb \,  =  \,\frac{1}{k^2  + q^2};
\eea
we obtain 
\beq \label{FPHI9}
\Phi_{\mbox{proton}}\Lb z\Rb\,=\, 2^{5 - 2\Delta}\,q^2\,z^{6 - 2\Delta}\,\,\int^{\infty}_0\frac{K^2_{1 - \Delta}(t)\,t^{2 \Delta - 1 } \,d t}{\Lb t^2  + q^2 z^2\Rb^2}
\eeq
\subsection{Physical observables}
We deal with the DID structure $F_2$ which can be written in the form
\beq \label{PO1}
F_2\Lb Q^2; x\Rb \,\,\,=\,\,\,\frac{Q^2}{4 \pi^2 \alpha_{em}}\,\,\sigma_{tot}\Lb \gamma^* + p ; \mbox{\eq{HES4}}\Rb
\eeq

For proton-proton collision we will use the total cross section  written as
\bea \label{PO2}
&&\sigma_{tot}\Lb p+ p \Rb\,\,=\,\, \int^\infty_0  d z_1 d z_2 \Phi_{\mbox{proton}}\Lb z_1\Rb\,\Phi_{\mbox{proton}}\Lb z_2\Rb\\\
&&\,\,2 \int d^2 b \,\,\Big\{1\,\,-\,\,\cos\Lb N^2_c\,\mbox{Re}\tilde{A}\Lb s,b;z_1,z_2\Rb\Rb\,\exp\Lb - N^2_c\,\mbox{Im}\tilde{A}\Lb s,b;z_1,z_2\Rb\Rb\Big\}\nonumber
\eea
with $\Phi_{\mbox{proton}}\Lb z\Rb$ given by \eq{FPHI9}. However \eq{HES1} describes the scattering amplitude only in the region of high energies. For DIS we select onlt data with $x \leq 0.02$ and we use for fitting the following expression:
\beq \label{PO3}
F_2\Lb Q^2; x\Rb \,\,\,=\,\,\,F_2\Lb Q^2; x; \mbox{\eq{PO1}}\Rb \,\,+\,\,F^{in}_2\Lb Q^2\Rb
\eeq
where $F^{in}_2\Lb  Q^2\Rb$ is a DIS structure function at $x_0 = 0.02$ and it  was considered as   a fitting parameter at any  value of $Q$.

For proton-proton interaction we described the data at $W  = \sqrt{s} \geq \,20 \,GeV$
and add a constant $\sigma_0$ to \eq{PO2}.

\eq{HES3}  which is written in N=4 SYM, has power - like decrease at large valies of the impact parameter ($b$). On the other hand for  not very small $\rho$ the conformal symmetry of N=4 SYM is broken and we need to consider the string theory
for which N=4 SYM is a conformal limit at small $\rho$. In the string theory the hadron spectrum has the lightest hadron with the mass $m_{\mbox{glueball}} \,=\,\,\sqrt{\rho/\alpha'}$ which leads to the exponential falldown of the amplitude $\tilde{A}\Lb s, b; z_1,z_2\Rb $ at large $b$: $\tilde{A}\Lb s, b; z_1, z_2\Rb \xrightarrow{b \gg m_{\mbox{glueball}]} } \,\exp\Lb - m_{\mbox{glueball}}\,b \Rb$. We rewrite \eq{HES3} in the form
\beq \label{PO4} 
\tilde{\cal A}\Lb s,b; z_1. z_2\Rb\,\,=\,\,\tilde{A}\Lb s,b; z_1,z_2|;\eq{HES1}\Rb\,e^{ - m_{\mbox{glueball}}\,b }
\eeq
to take into account the mass spectrum of the string theory. The final answer for the amplitude in this case is \eq{HES3} in which
$\tilde{A}$ is replaced by $\tilde{A}$,  ($ \tilde{A} \rightarrow \tilde{\cal A}$).

In this paper we check also our description of the DIS data with the experimental data on the total cross section for proton-proton scattering. For this observable we use the following expression:
\beq \label{PO5}
\sigma_{\mbox{ proton - proton }}\Lb s\Rb\,\,=\,\,\sigma_{\mbox{proton-proton}}\Lb \eq{PO2}\Rb \,\,+\,\,\sigma_0\Lb s \Rb
\,\,\,\,\,\,\mbox{with}\,\,\,\,\,\sigma_{0}\Lb s\Rb \,=\,\,\sigma_{01}\,\,+\,\,\frac{\sigma_{02}}{\sqrt{s}}
\eeq
The contribution $\propto 1/\sqrt{s}$ corresponds to the contribution of the secondary Regge poles. $\sigma_{0}\Lb s \Rb$ is related to the mechanism of the strong interaction that cannot be described by N=4 SYM or  to unknown corrections to this theory
 $\propto 1/\lambda$.
\section{Comparison with the experimental data}

\TABLE{
\begin{tabular}{|l|l|l|l|l|}
\hline \hline
Solution & $\rho$ & $g$& $\Delta$ & $\chi^2/d.o.f.$\\
\hline \hline
I & 0.701 $\pm$ 0.004 & 0.004 $\pm$ 0.001 & 2 &1.42\\
\hline
II& 0.75 $\pm$ 0.007 &4.019 $\pm$ 0.061 & 2& 1.22\\
\hline \hline
\end{tabular}
\caption{Fitting parameters for solution I (see \eq{HES1}) and solution II (see \eq{PO4}).}
}

Using the formulae of the previous section we  compare the value of $F_2\Lb Q^2;x\Rb$ with the HERA experimental data  in the region of  low $x$ ( $x\,\leq\,0.02$) .  As has been  mentioned our main goal is   to obtain two parameters of the N=4 SYM: $\rho$ and $g_s$.  For each chosen value of $Q^2$ we introduce one more phenomenological parameter: the value of $F_2(Q^2)$.
It should be mentioned that the value of $\Delta$ in \eq{FPHI3}  for the proton  as well as the value of $q$ have to be found from the fit, but we have to recall that the value of $q$  characterizes the typical scale of the non-perturbative wave function of the proton and can be extracted from the electromagnetic radius of the proton.

As we have mentioned the main qualitative experimental observation is that $F_2 \propto (1/x)^{\lambda(Q^2)}$ and the power $
\lambda(Q^2)$ depends on $Q^2$ changing from $\lambda \approx 0.1\div 0.2$ at low $Q^2 \leq 1\,GeV^2$ to $\lambda \approx 0.4$  at high $ Q^2 > 50\, GeV^2$ (see \fig{slopef2}).

 At first sight we cannot reproduce such a behaviour
since the exchange of the Pomeron generates only one power $ \lambda = 1 -  \rho$. The  only way out is to include the Pomeron re-scattering which could lead to the amplitude with the effective power that depends on $Q$. To our surprise we fitted the HERA data with small value of $g$ which turns out to be so small that in the accessible region of energies including the LHC highest energy
the Pomeron re-scattering does not contribute and only the exchange of the one Pomeron determines the amplitude. The restoration of the unitarity constraint will occur at ultra high energy.    

\FIGURE[ht]{
  \includegraphics[width=10 cm] {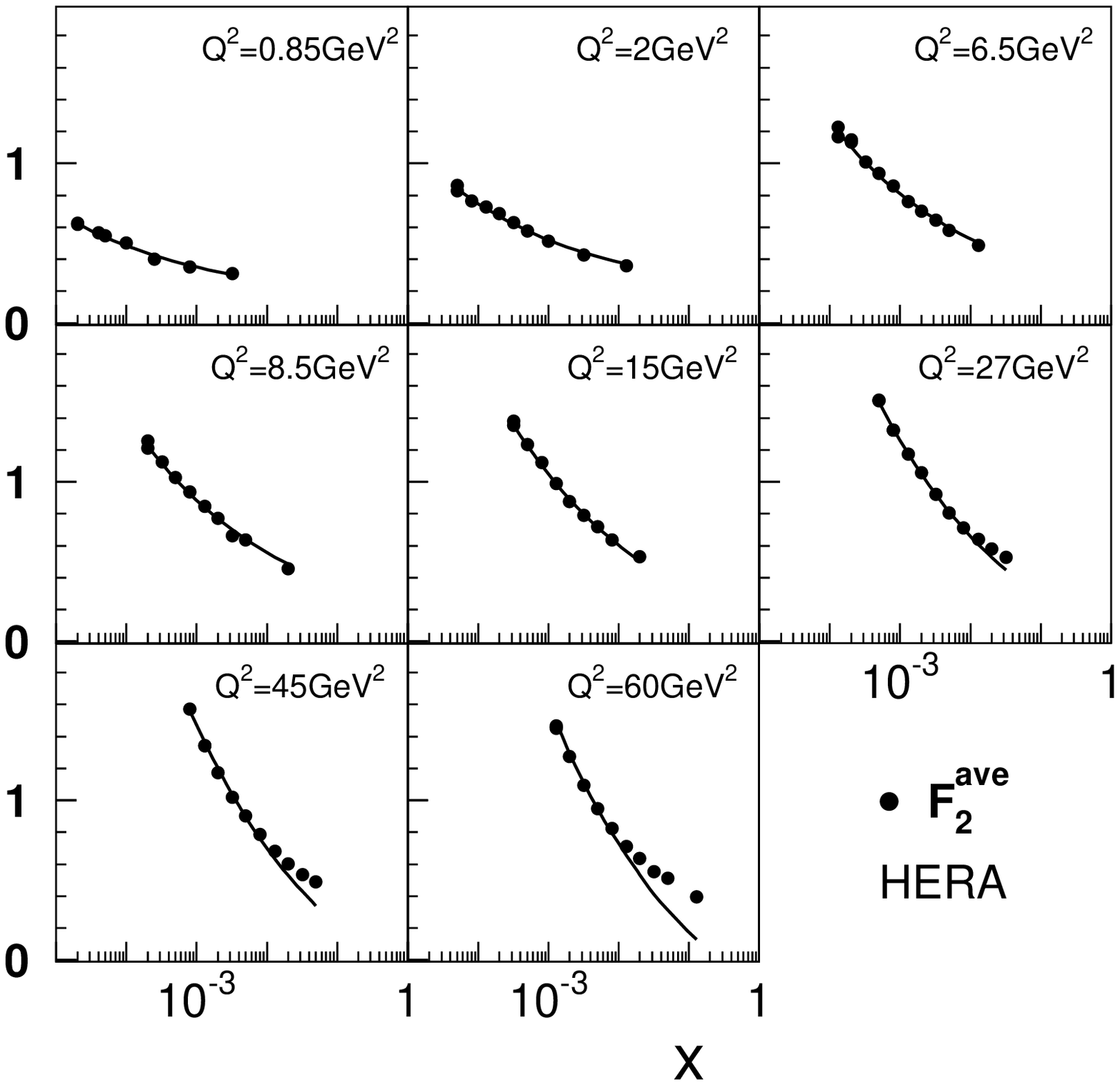}
\caption{The new HERA data on $F_2$ versus $x$ at fixed $Q$\cite{HERANEWDATA}. $ \rho = 0.75, g_s= 4.019,\Delta = 2$. }
\label{f2fit}
}

~

The quality of description one can see from \fig{f2fit} where the new HERA data\cite{HERANEWDATA} are plotted at eight values of $Q^2$ as function of $x$. The message is clear: the $z$ and $\ln s $ dependence in \eq{HES1} reproduces the
change in $\lambda$ as function of $Q$ shown in \fig{slopef2}. 

The fit to the experimental data was made  in two differrent cases: the first one ( solution I) corresponds to \eq{HES3} with the  amplitude $\tilde{A}$ determined by \eq{HES1}; and the second takes into account \eq{PO4} which restricts the integral over $b$ (solution II).  The parameters that we obtain are listed in Table 1. Solution I gives small value of
 $g = N^2_c g_s= 0.04$  while solution II leads to rather large  $g = N^2_c g_s= 4$. They have very close $\chi^2/d.o.f. \approx 1.5$ and describe the data equally well (see \fig{2sol}). However both solutions cannot describe the data at lower $Q$ (see \fig{2sol}-c and \fig{2sol}-d).  From the fit we also determine the function $F^{in}_2$ in \eq{PO3} ( see \fig{inf2}).

\FIGURE[t]{\begin{tabular}{c c c  }
  \includegraphics[width=4.5cm] {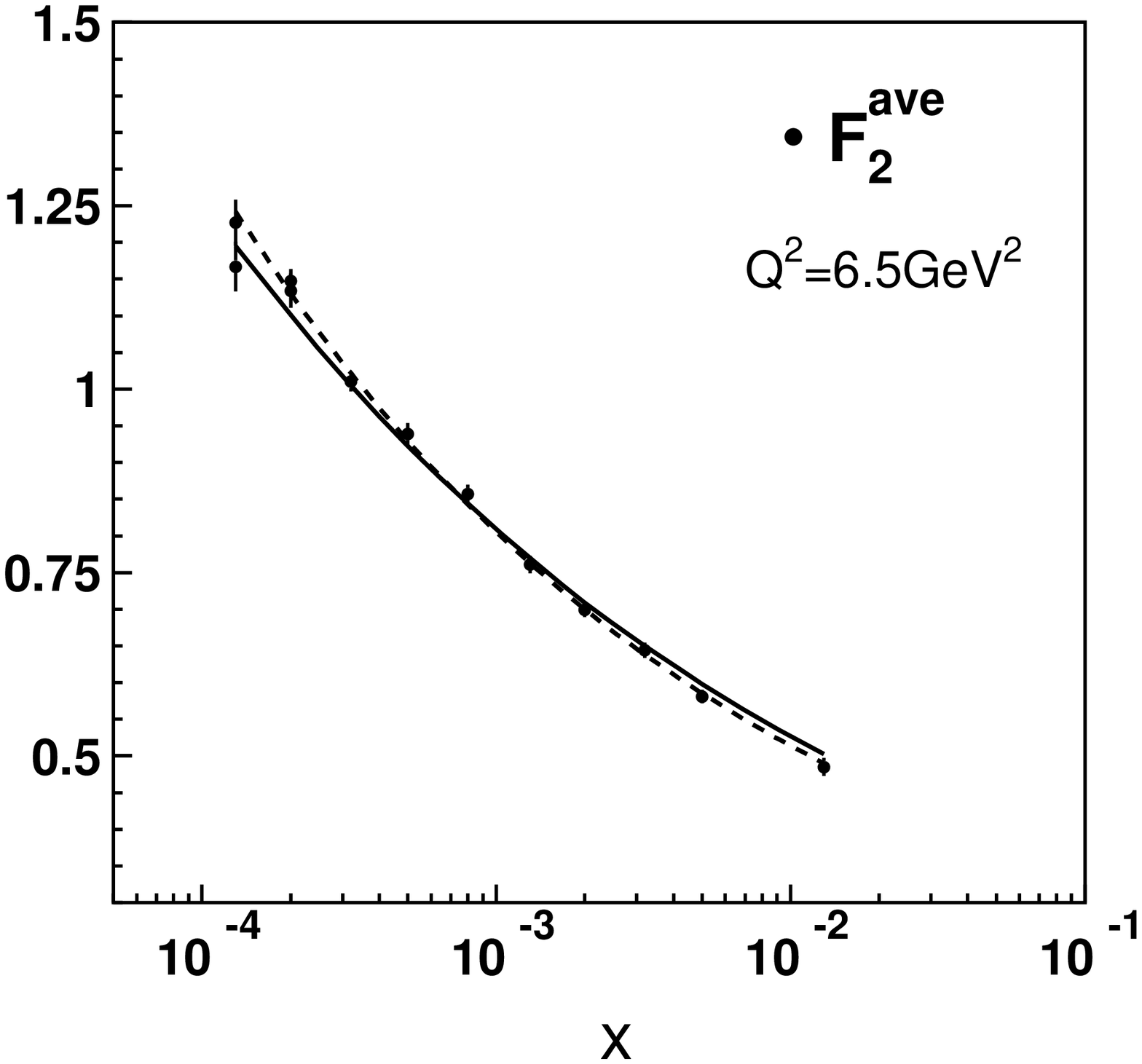} &  \includegraphics[width=4.5cm] {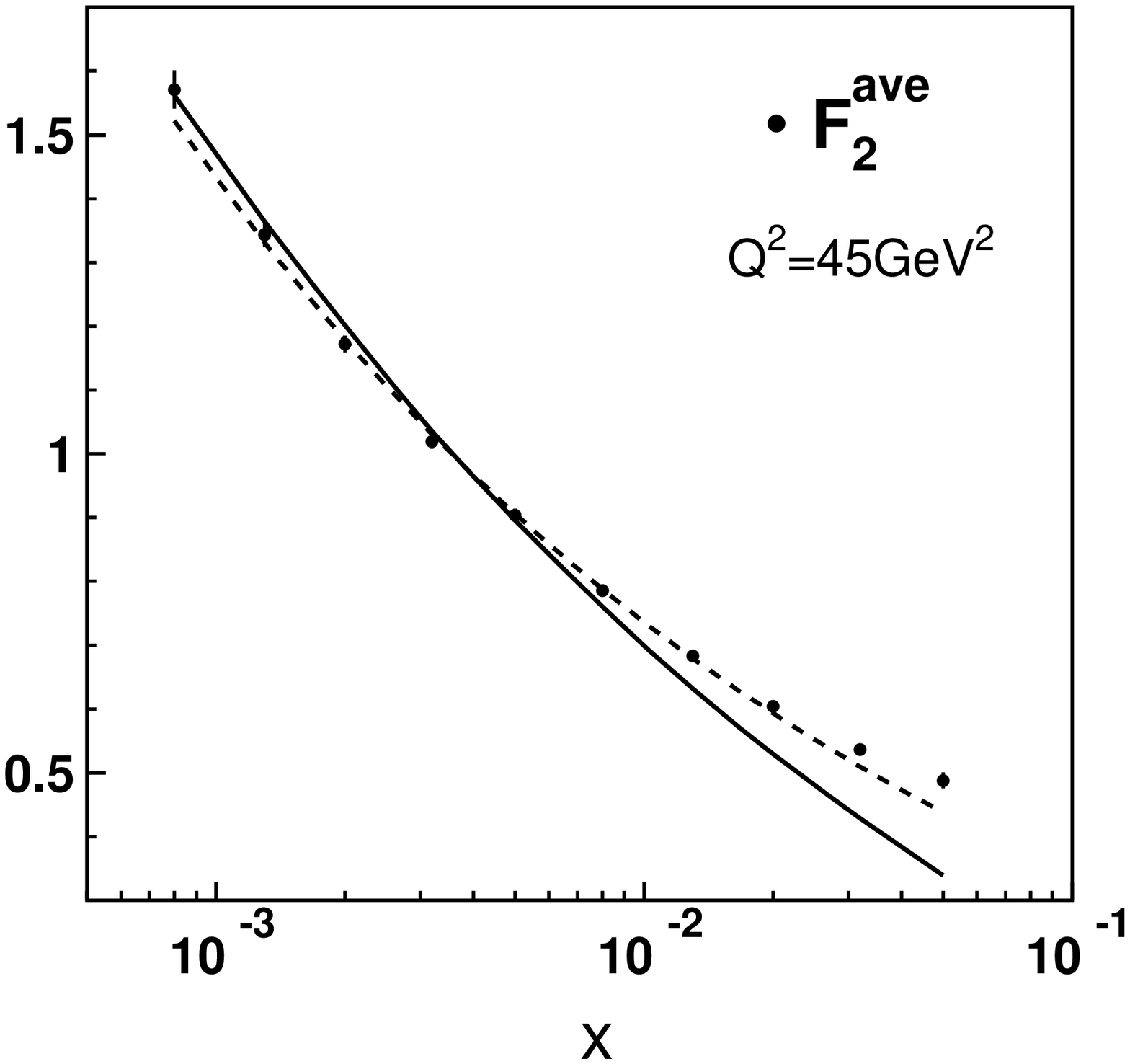}&  \includegraphics[width=4.5cm] {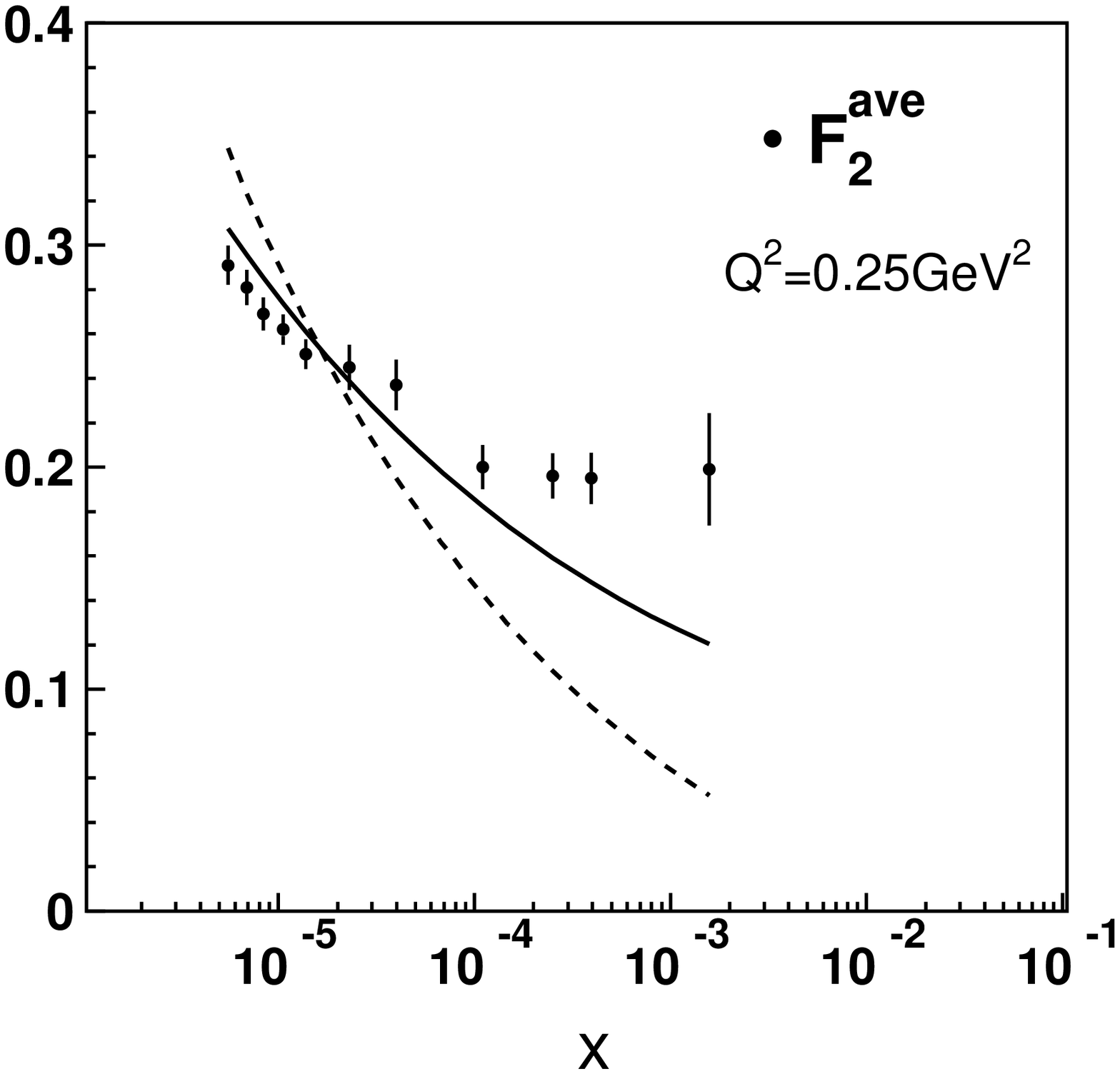} \\
~&~&\\
\fig{2sol}-a & \fig{2sol}-b &\fig{2sol}-c\\
\end{tabular}
\caption{ Comparison of
solution I (dotted line) and solution II ( solid line)
for three values of $ Q^2$
(all other information is in the pictures.)}
\label{2sol}}

It turns out that for both solutions the amplitude $A$ is small ( see \fig{amp} that illustrated this fact).
In this figure one sees that the amplitude $\tilde{\cal A}$ for the solution II reaches the value of about 0.5 but in spite the fact that this value does not look very small \eq{HES3} gives the value of the amplitude which is very close to $\mbox{Im} \tilde{\cal A}$. For the solution I the value of $g$ is so small that it leads to  a very small  $\tilde{A}$. Therefore, the lesson which we obtain from this estimates is very simple: the data for DIS can be described in N=4 SYM but
 the non-linear (shadowing) corrections tuns out to be very small. In other words the saturation effects which are in N=4 SYM are very similar to the one in QCD\cite{MHI,BST1,BST2,BST3,COCO,LMKS}, will be sizeable only at  ultra high energy, higher than the LHC maximum energy ($W = 14 \,TeV$).

These two solutions we
check against the experimental data for the total cross section of proton-proton interactions. The comparison with the experimental data is shown in \fig{sig}

\EPSFIGURE[b]{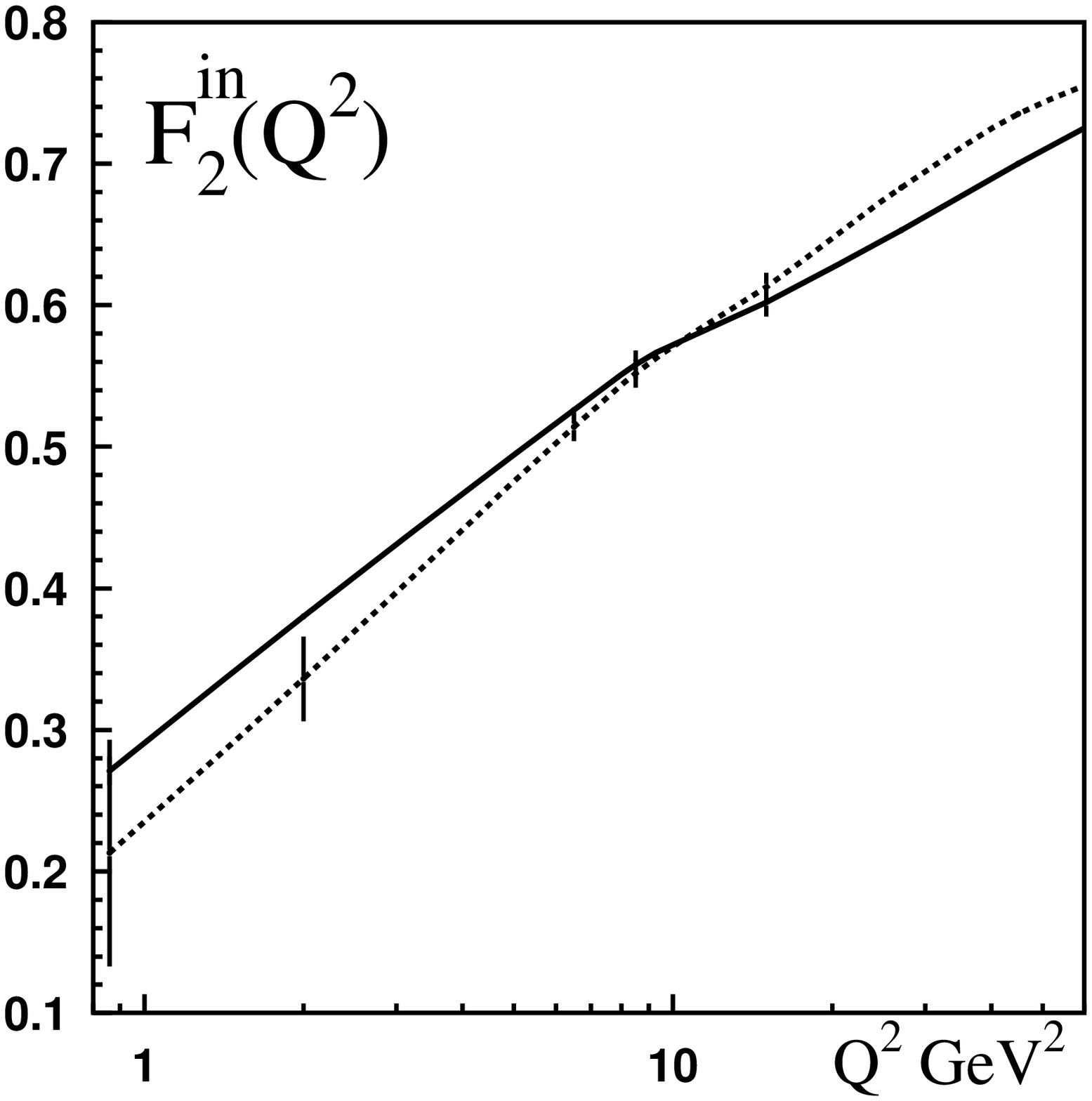,width=6cm}{$F_2\Lb x = 0.01; Q^2\Rb \equiv F^{in}_2\Lb Q^2\Rb$ versus $Q^2$ for solution I(dotted line) and solution II(solid line).
In this figure the errors are shown since the values of $F^{in}_2\Lb Q^2\Rb$ were considered as independent fitting parameter.\label{inf2}} 

~

~

~

One can see that with $\sigma_0$ which does not increase  with energy we cannot obtain the good fit of the experimental data for the total proton-proton cross section. However, it is clear that the solution II  gives the description closer to the data in comparison with the solution I. The fact that we did not obtain a good description of the data for the proton-proton scattering does not look discouraging to us since we made oversimplified assumption about $\Phi_{\mbox{proton}}$: colourless dipoles  are  correct d.o.f. at high energy and $K_0(r q)$ is the wave function of the dipole.  It should be stressed that the unknown mechanism which is different from N=4 SYM and which leads to $\sigma_0$ contribution in \eq{PO5} is responsible only for the half of the total inelastic cross section at RHIC energies ($W = 300\,GeV$) and less than a quarter for the LHC energies.

\EPSFIGURE[b]{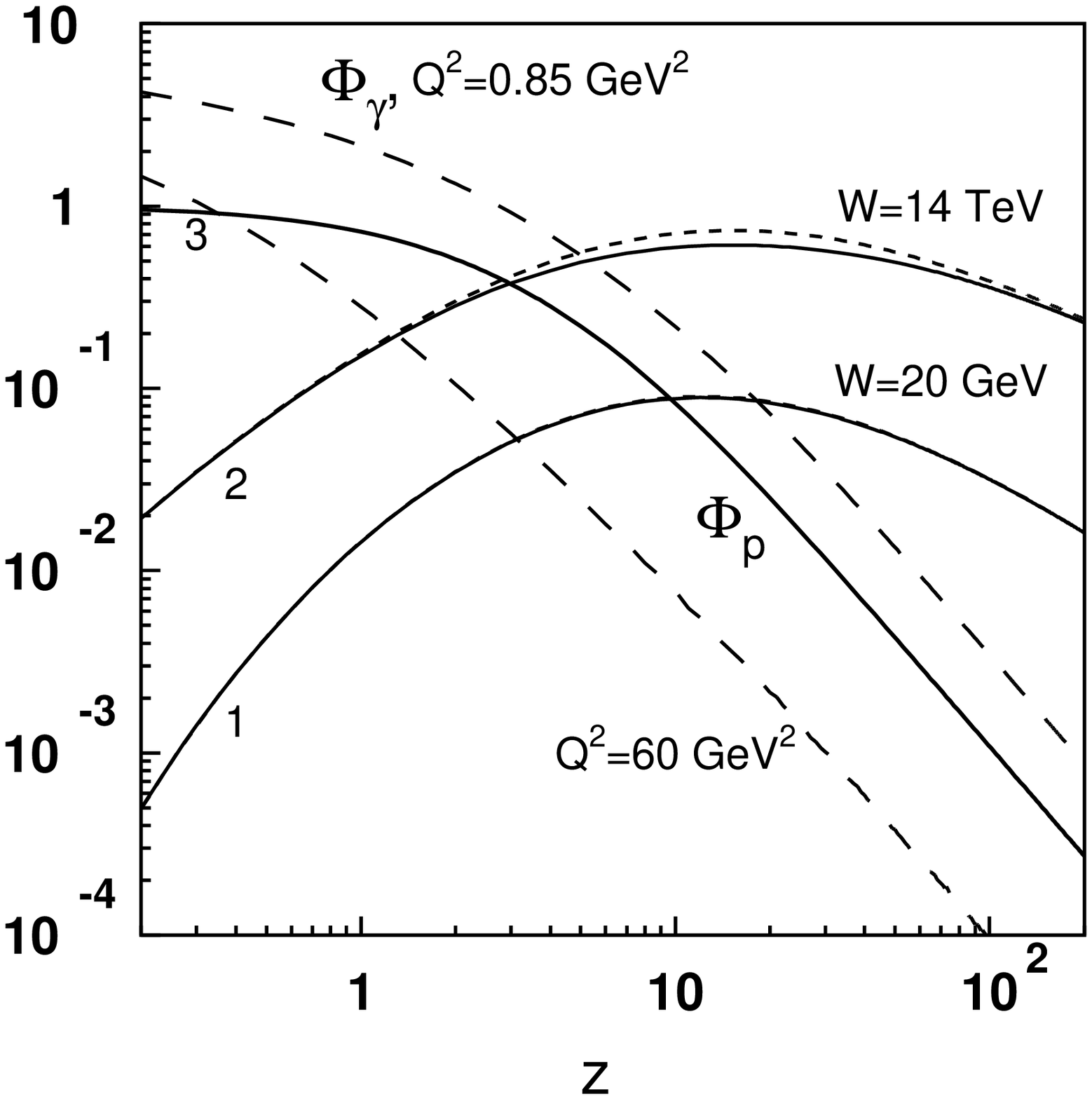,width=6cm}{Functions  $\Phi_{\gamma}$ (two dashed lines), $\Phi_{\mbox{proton}}$ (solid line 3) and the amplitude of 
\protect\eq{HES3} (solid lines 1 and 2)  as functions of $z_2 = z$ for fixed $z_1 = 10 GeV^{-1}$  at two energies $W = \sqrt{s} = 20 \,GeV$ and $W = \sqrt{s} =14 \,TeV$.
Dotted curve presents $\mbox{Im}{\cal A} $ of \protect\eq{PO4} at $W = \sqrt{s} =14 \,TeV$.
\label{amp}
}

~
\section{Conclusions}

As has been mentioned comparing the N=4 SYM prediction with the experimental data we obtain two surprising results.
First,  the N=4 SYM formula gives a good description of the DOIS structure function in wide range of $Q^2$ ($ Q^2 = 0.85 \div 60\,GeV^2$ and $x$ ( $x \leq 0.01$). The surprise stems from the fact that \eq{HES1} leads to power-loke dependence on $x$ ($F_2 \propto (1/x)^{1 - \rho}$ ) and this power does not depend on $Q$, Comparing this formula with \fig{slopef2} one can conclude that dependence of $z$ of \eq{HES1} as well as on $\ln s$ simulates the effective power dependance on $Q$.

The second surprise is the smallness of $g_s$ that generates a very small shadowing corrections which we can neglect even at the highest accessible energy:$W = 14\,TeV$.  $\rho = 0.7 \div 0.75$ means that the intercept of the Pomeron is rather small $\Delta_{\pom} \approx 0.3 \div 0.35$. We used to consider such a small intercept to be  typical for the weak coupling limit (for the BFKL Pomeron). On the other hand, in small coupling limit we expect a strong shadowing correction induced by the Pomeron interactions ( see Refs. \cite{GLR,MUQI,BK,JIMWLK}).  Recalling that $\rho = 0.7 \div 0.75$ corresponds  to $\lambda  \approx 7 \div 8$ we could expect that the corrections of the order of $1/\lambda^2$ will be small. In this case we expect the small shadowing corrections with our fitted small value of $g_s$.  Therefore, we have a dilemma: either the corrections of the order of $1/\lambda^2$  are large or the shadowing phenomenon is neglidgibly small.

The influence of the corrected $b$ dependence was expected but the fact that even with corrected $b$ dependence we have still small shadowing corrections was not expected.

In general we believe this analysis of the experimental data in the framework of N=4 SYM theory gives the useful information on the possible scenario what is going on in strong coupling limit at high energy. The picture that arises from this analysis is in clear contradiction from the expectation of the high density QCD and because of this it could lead to a better understanding the matching between soft (large coupling) and hard (small coupling) processes in QCD.

\newpage

\section* {Acknowledgements}

\FIGURE[b]{\includegraphics[width=6cm]{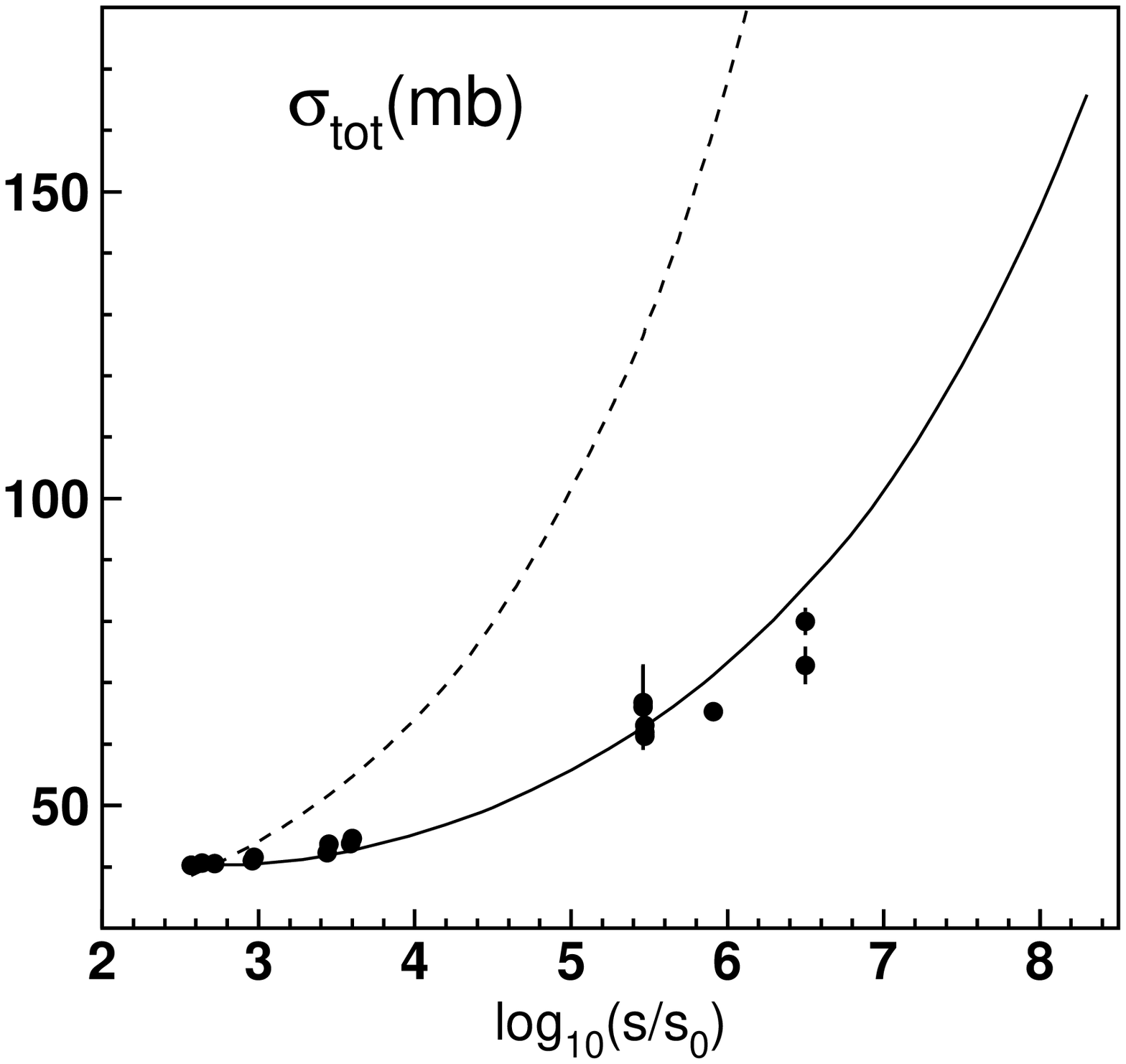}
\caption{The comparison of the total cross section for proton-proton high energy scattering with N=4 SYM predictions for solutions I (dotted line) and II (solid line).}
\label{sig}
}

~

~

~

~

We thank Boris Kopeliovich and Chung-I Tan for fruitful discussions on the subject of this paper. Chung -I Tan was the first who mentioned to E.L. in private discussion that the description of the DIS data could lead to a very small $g_s$. Actually this remark was the main impetus for this paper.

 This work was supported in part by Fondecyt (Chile) grants, numbers 1090236  and 1100648.

\end{document}